\documentstyle[12pt]{article}
\input epsf.tex
\def\beq{\begin{equation}}
\def\eeq{\end{equation}}
\def\eeq{\end{equation}}
\def\bea{\begin{eqnarray}}
\def\eea{\end{eqnarray}}
\def\bq{\begin{quote}}
\def\eq{\end{quote}}

\def\dsl{\not\!\partial}
\def\CS{{\cal S}}
\def\bn{\left|L\,n\right\rangle}
\def\fn{\left|R\,n\right\rangle}
\def\bnd{\left\langle L\,n\right|}
\def\fnd{\left\langle R\,n\right|}

\def\ket#1{\left|#1\right\rangle}
\makeatletter
\def\vereq#1#2{\lower3pt\vbox{\baselineskip1.5pt \lineskip1.5pt
\ialign{$\m@th#1\hfill##\hfil$\crcr#2\crcr\sim\crcr}}}
\makeatother

\title{Hierarchies without Symmetries from Extra \\ Dimensions}
 
\author{Nima Arkani-Hamed and Martin Schmaltz\ \thanks{\tt nima@slac.stanford.edu, schmaltz@slac.stanford.edu}\\ \\
  \small \sl SLAC, Stanford University, Stanford, CA 94309\\ \\
  }

\begin{document}
\baselineskip=17pt
\pagestyle{plain}

\begin{titlepage}
\maketitle
\begin{picture}(0,0)(0,0)
\put(295,300){SLAC-PUB-8082}
\end{picture}

\begin{abstract}
\leftskip-.6in
\rightskip-.6in
\vskip.4in
It is commonly thought that small couplings in a low-energy
theory, such as those needed  for the fermion mass hierarchy
or proton stability, must originate from symmetries in a
high-energy theory. We show that this expectation is violated
in theories where the Standard Model fields are confined to a
thick wall in extra dimensions, with the
fermions ``stuck" at different points in the wall.
Couplings between them are then suppressed due to the exponentially
small overlaps of their wave functions. This provides a framework
for understanding both the fermion mass hierarchy and proton
stability without imposing symmetries, but rather in terms
of higher dimensional geography.  
A model independent prediction of this scenario is non-universal
couplings of the Standard Model fermions to the ``Kaluza-Klein''
excitations of the gauge fields. This allows a measurement
of the fermion locations in the extra dimensions at the LHC or NLC
if the wall thickness is close to the TeV scale.  
\end{abstract}
\thispagestyle{empty}
\setcounter{page}{0}
\end{titlepage}

\section{Introduction}
The usual way of organizing our thinking about physics beyond the
Standard Model (SM)
is the effective field theory paradigm: all operators consistent
with the symmetries are present in the theory, 
with higher-dimension operators suppressed by powers of the
ultraviolet cutoff. The SM itself provides an 
exception to this expectation: the Yukawa couplings for all the
fermions other than the top quark are
much smaller than ${\cal O}(1)$. This does not lead to any
fine-tuning problems since small Yukawa couplings are
technically natural. Nevertheless, we are normally led to suspect that 
the fermion mass hierarchy is controlled by (weakly broken) flavor
symmetries operative at shorter distances.
Similar issues surround the question of proton decay in extensions
submof the SM, especially when there is new physics at the TeV scale. 
Once again, some symmetry is normally invoked to forbid dangerous
1/(TeV) suppressed interactions
mediating proton decay. Furthermore, imposing global symmetries on
low-energy effective theories, for instance,
stabilizing the proton by declaring that the low-energy theory
respects baryon number, is widely considered to be unsatisfactory
given the lore that black-holes/wormholes violate all non-gauged
symmetries. This seems particularly problematic 
for theories where the fundamental Planck scale is lowered close
to the TeV scale \cite{ADD,AADD,ADD2,ADHMR,other,DDG},
and suggests that some sort of 
continuous \cite{ADD2,AD} or discrete \cite{AADD} 
gauge symmetry is required to adequately suppress proton decay.
 
In this paper, we will show that all of this lore can easily
and generically be violated in theories where the 
SM fields are constrained to live on a wall in $n$ extra dimensions,
where gravity and perhaps other SM singlet fields
are free to propagate. We will construct a simple model where our
wall is slightly thick in one of the extra dimensions.
The wall will have interesting sub-structure: while the Higgs and
SM gauge fields are free to propagate inside it, the SM fermion are
``stuck" at different points in the wall, with wave functions given
by narrow Gaussians as shown in Figure 1. 

\begin{figure}[ht]
\centering
\epsfxsize=5.0in\epsfbox{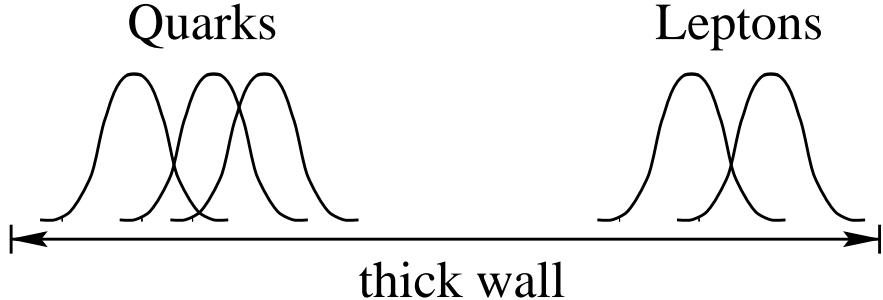}
\smallskip
\parbox{5in}{\caption{\it Profile of Standard Model fermion wave
functions (vertical axis) in the extra dimensions (horizontal axis).
The fermions freely propagate in 3+1 dimensions
(not shown) and are ``stuck''
at different locations in the extra dimensions. The gauge and
Higgs fields' wave functions occupy the whole width of the
thick wall. Direct couplings between the fermions are then suppressed
by the exponentially small overlap of their wave functions. If --
as shown here -- quarks
and leptons live on opposite ends of the wall profile protons
become essentially stable. The hierarchy of Yukawa couplings
arises from order one (in units of the fermion wave function
width) distances between left and right handed
components of the fermions.
}}
\end{figure}

Without imposing any flavor symmetries on the short-distance theory,
we will see that the long-distance 4-dimensional theory
can naturally have exponentially small Yukawa couplings, arising
from the small overlap between  
left- and right-handed fermion wave functions. Similarly, without
imposing any symmetries to protect against proton decay,
the proton decay rate can be exponentially suppressed to safety
if the quarks and leptons are are localized at different ends
of the wall \footnote{Our approach to 
to the fermion mass hierarchy similar in spirit to
the one in \cite{AD}. For other approaches to 
suppressing Yukawa couplings and proton decay, see \cite{DDG}.}.
We emphasize that there is nothing fine-tuned about this from the
point of view of the low-energy 4-dimensional theory; all the exponentially
small couplings are technically natural. However, our examples violate
the usual intuition that small couplings in a low-energy theory
must be explained by symmetries in the high-energy theory. Instead,
small couplings arise from the location and geometry
of fermion fields stuck at different points in the extra dimensions,
with no symmetries in the high-energy theory whatsoever. Note that this
mechanism of separating fermions in an extra dimension is already
being used to preserve chiral symmetry on the lattice in
Kaplan's domain wall fermions \cite{david}.
Lattice simulations \cite{latt} show that chiral symmetry is
protected very effectively by separating the left and right
handed components of the fermions in the 5'th dimension.

If the wall thickness $L$ is close to the TeV scale, which is natural in
theories with very low fundamental Planck scale, 
the mechanisms suggested in this paper can give rise to dramatic
signals at future colliders.  Since the SM gauge fields can only
propagate inside 
the wall, $L$ effectively acts as the size of the extra dimensions for
them\footnote{Note that the dimensions where the gauge fields propagate
need not be orthogonal to the large dimensions in which only gravity
propagates; the gauge fields can just be restricted to live in a
smaller part of the gravitational dimensions. The possibility of TeV sized
extra dimensions with KK excitations for the SM gauge fields was first considered by 
Antoniadis \cite{Ignatios}}.
Therefore, at energies above $L^{-1}$, ``Kaluza-Klein"
excitations  (the higher harmonics of a particle in a box) of the gauge fields
can be produced, and can scan the wall substructure.
In particular, while the lowest
excitation of the gauge fields (which we identify as the usual 4-d
SM gauge fields) have a flat wave function throughout the wall and
couple with standard strength to all the SM fermions, the KK
excitations couple with non-universal strength
to the fermions stuck at different points in the wall. For instance,
if some of the fermions are stuck at special points (say the center
of the wall),
KK excitations of e.g. the photon can be baryophobic or leptophobic.
More generally, measurements of the non-universal couplings of KK
excitations to SM fermions can pin down their geometrical
arrangement in the thick wall.

We emphasize that our prediction of
non-universal couplings of the SM fermions to gauge and Higgs fields
is model-{\it independent}, it only depends on the fact that the
fermions are stuck at different points in the extra dimensions. Of
course, the values of the different couplings are model-{\it dependent} and
can be used to distinguish between models.

In Section 2 we describe an explicit field theory mechanism
which we use to construct a setup as outlined above; we discuss
how to localize a single chiral fermion to defects in higher dimensions
and then generalize to several fermions localized at different points
in the vicinity of the same defect. In Section 3 we derive the
exponentially small couplings which result from our framework and
demonstrate how the scenario can explain the SM fermion mass hierarchy
and suppress proton decay. We also comment on neutrino masses.
Section 4 contains a brief discussion of experimental signatures
resulting from the non-universal couplings of KK gauge fields. For example,
our KK fields make a contribution to atomic parity violation with the
correct sign to explain the discrepancies between the SM prediction
and the most recent experimental results \cite{apv}.
Our conclusions are drawn in section 5.

\section{Localizing chiral fermions}
\subsection{One chiral fermion in 5 dimensions}
For simplicity we limit ourselves to constructions with one extra
dimension. Generalizations to higher dimensions are equally interesting
and can be analyzed similarly. Localizing fields in the extra dimension
necessitates breaking of higher dimensional translation invariance.
This is accomplished in our construction of a thick wall by a
spatially varying expectation value for a five-dimensional scalar
field $\Phi$ as shown in Figure 2. We assume the expectation
value to have the shape of a domain wall transverse to the extra
dimension and centered at $x_5=0$. 
For example, such an expectation value
could result from a ${\bf Z}_2$ symmetric potential for $\Phi$.
\footnote{Interactions with the fermions below break this symmetry
and render the domain wall profile unstable but the rate for
tunneling to a constant expectation value can easily be suppressed
to safety.}

\begin{figure}[ht]
\centering
\epsfxsize=5.0in\epsfbox{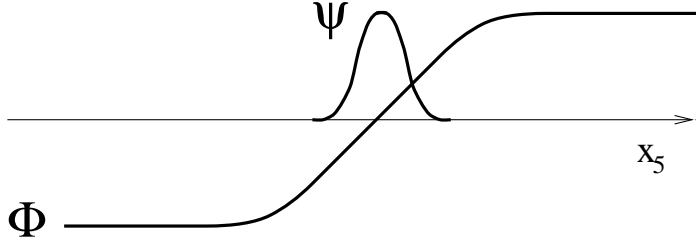}
\smallskip
\parbox{5in}{\caption{\it Profile of the scalar domain wall field
$\Phi$ in the $x_5$ dimension. A chiral zero mode fermion is localized
at the zero of $\Phi$.
}}
\end{figure}

We will now show that the Dirac equation for a five dimensional
fermion in the background of this scalar field has a zero mode
solution which corresponds to a four dimensional chiral fermion
stuck at the zero of $\Phi$ \cite{Jackiw}. A convenient representation for the
$4 \times 4$ gamma matrices in five dimensions is

\beq
\gamma^i = \left( \begin{array}{cc}
                    0 & \sigma^i \\
                    \overline{\sigma}^i & 0
                  \end{array} \right)\ , \quad i=0..3\ , \quad
\gamma^5= -i\ \left( \begin{array}{cc}
                    {\bf 1} & 0 \\
                    0 & {\bf -1}
                  \end{array} \right) .
\label{gammas}
\eeq
As it will be useful in the following sections, we record below the
two different Lorentz invariant fermion bilinears in 
5 dimensions
\beq
\bar\Psi_1 \Psi_2 , \quad \Psi^{T}_1 C_5 \Psi_2 
\eeq
where
\beq
 C_5 = \gamma^0 \gamma^2 \gamma^5 = \left(\begin{array}{cc}
\epsilon & 0 \\ 0 & -\epsilon \end{array} \right) \quad
\mbox{in the Weyl basis}.
\label{C5}
\eeq
The first is the usual Dirac bilinear, while the second is the
Majorana bilinear which generalizes the familiar 4-dimensional expression,
where instead of $C_5$ we have $C_4 = \gamma^0 \gamma^2$.

The action for a five dimensional fermion $\Psi$ coupled to the background
scalar $\Phi$ is then
\beq
\CS = \int {\rm d}^4{\bf x}\ {\rm d}x_5\,
\overline\Psi[i\!\!\dsl_4+i\gamma^5\partial_5+\Phi(x_5)] 
\Psi\ .
\label{single5}
\eeq

Here the coordinates of our $3+1$ dimensions are represented by
${\bf x}$ whereas the fifth coordinate is $x_5$; five-dimensional
fields are denoted with capital letters whereas
four-dimensional fields will be lower case.
This Dirac operator is separable, and it is convenient to expand
the $\Psi$ fields in a product basis
\beq
\Psi({\bf x},x_5) = \sum_n \langle x_5 \bn P_L \psi_n({\bf x})
           + \sum_n \langle x_5 \fn P_R\psi_n({\bf x})  \nonumber \\
\eeq
\beq
\bar\Psi({\bf x},x_5) = \sum_n \bar\psi_n({\bf x}) P_R \bnd x_5 \rangle
               + \sum_n \bar\psi_n({\bf x}) P_L \fnd x_5 \rangle \ ,
\label{expansion}
\eeq
where the $\psi_n$ are arbitrary four-dimensional Dirac spinors
and $P_{L,R}=(1\pm i\gamma^5)/2$ are chiral projection operators.
We use a bra-ket notation for the eigenfunctions
which diagonalize the $x_5$-dependent part of the Dirac operator;
the kets $\bn$ and $\fn$ are solutions of 
\bea
\label{susyqm}
aa^\dagger \bn=(-\partial_5^2+\Phi^2+\dot\Phi) &\bn =\mu_n^2 \bn
       \nonumber \\
a^\dagger a \fn=(-\partial_5^2+\Phi^2-\dot\Phi) &\fn=\mu_n^2 \fn \ ,
\eea
respectively. Here $\dot\Phi\equiv\partial_5\Phi$, and $a^\dagger$ and $a$
are ``creation'' and ``annihilation'' operators defined as
\bea
a=&\partial_5+\Phi(x_5) \, \nonumber \\
a^\dagger=&-\partial_5+\Phi(x_5) \ .
\eea
The $\bn$ and $\fn$ each form an orthonormal set and for non-zero
$\mu_n^2$ are related through
$\fn = (1/\mu_n)\, a\, \bn$ as can
be verified easily from Eq.(\ref{susyqm}). The eigenfunctions with vanishing
eigenvalues need not be paired however. 
It is no accident that we use simple harmonic oscillator (SHO) notation.
For the special choice $\Phi(x_5)=2 \mu^2 x_5$ the operators $a$ and
$a^\dagger$ become the usual SHO creation and annihilation operators
up to a normalization factor $\sqrt{2} \mu$, and the operator
$a^\dagger a$ becomes the number operator $N$. The eigenkets are
then related to the usual SHO kets by $\bn=\ket{n}$ and $\fn=\ket{n-1}$.

The pairing of eigenfunctions also persists for general $\Phi$. This
follows most elegantly from considering the operators $Q=a\gamma^0 P_L$
and $Q^\dagger=a^\dagger\gamma^0 P_R$ which are the supercharges
of an auxiliary supersymmetric quantum mechanics system
\cite{wittensusyqm} with Hamiltonian
$H=\{Q,Q^\dagger\}$. Then $P_L \bn$ and $P_R \fn$ are the ``boson''
and ``fermion'' eigenstates of $H$ respectively, and the equality
of eigenvalues of $\bn$ and $\fn$ is the usual boson-fermion degeneracy
of supersymmetric theories. Again, zero modes need not be paired which
allows us to obtain chiral 4-d theories.
While most of what follows applies also to the case of general
$\Phi$ we will find it convenient to use the SHO language.

Expanding in $\bn$ and $\fn$ the action for a 5-d Dirac fermion
eq. (\ref{single5}) can be re-written in terms of a 4-d action
for an infinite number of fermions
\beq
S = \int {\rm d}^4 {\bf x} \left[
\bar\psi_L\ i\!\!\dsl_4 P_L\ \psi_L +
\bar\psi_R\ i\!\!\dsl_4 P_R\ \psi_R +
\sum_{n=1}^{\infty} \bar\psi_n (i\!\!\dsl_4 +\mu_n) \psi_n   \right] \ .
\eeq
The first two terms correspond to 4-d two-component chiral fermions,they arise
from the zero modes of Eq.(\ref{susyqm}).
The third term describes an infinite tower of Dirac fermions corresponding
to the modes with non-zero $\mu_n$ in the expansion.

The zero mode wave functions are easily found by integrating
$a^\dagger \bn=0$ and $a \fn=0$.
The solutions
\beq 
\langle x_5 \ket{L,0}\sim{\rm exp}\left[-\int^{x_5}_0 \Phi(s)
{\rm d}s\right]\quad {\rm and}\quad
\langle x_5 \ket{R,0}\sim{\rm exp}\left[\int^{x_5}_0 \Phi(s)
{\rm d}s\right]\ ,
\eeq 
are exponentials with support near the zeros of $\Phi$.
In the infinite system that we are considering these modes  
cannot both be normalizable\footnote{Of course, we will be working in
finite volume in the end, then the other mode is normalizable as well,
but it is localized at the other end of the extra dimension. The existence
of this other mode is dependent on boundary conditions.}.
It is easy to see that $\ket{b,0}$ is normalizable
if $\Phi(-\infty)<0$ and $\Phi(+\infty)>0$ as in Figure 2, and if
$\Phi(-\infty)>0$ and $\Phi(+\infty)<0$ then
the mode $\ket{f,0}$ is normalizable. In the other cases there is no
normalizable zero mode.
For definiteness let us now specialize to the SHO. Then
\beq
\langle x_5 \ket{L,0}=\frac{\mu^{1/2}}{(\pi/2)^{1/4}}\ 
{\rm exp}\left[{-\mu^2{x_5}^2}\right]\ ,
\eeq
and $\langle x_5 \ket{R,0}$ is not normalizable. Thus the spectrum of
four dimensional fields contains one left-handed chiral fermion
in addition to an infinite tower of massive Dirac fermions. The
shape of the wave function of the chiral fermion is Gaussian,
centered at $x_5=0$.  
Note that coupling $\Psi$ to $-\Phi$ would have rendered
$\langle x_5 \ket{R,0}$
normalizable and we would have instead localized a massless right handed
chiral fermion.

For clarity, let us write the full wave function of the massless chiral
fermion in the chiral basis
\beq
\Psi({\bf x},x_5) = \left(\begin{array}{c} \langle x_5 \ket{L,0}
\psi({\bf x}) \\ 0 \end{array} \right)\ .
\eeq

\subsection{Many chiral fermions}

We can easily generalize Eq. (\ref{single5}) to the case of several
fermion fields. We simply couple all 5-d Dirac fields to the same
scalar $\Phi$

\beq
\CS = \int {\rm d}^5x\, \sum_{i,j} \bar\Psi_i[i\!\dsl_5+ \lambda\Phi(x_5)-m]_{ij} \Psi_j\ .
\label{multi5}
\eeq
Here we allowed for general Yukawa couplings $\lambda_{ij}$ and
also included masses $m_{ij}$ for the fermion fields. Mass terms
for the five-dimensional fields are allowed by all the symmetries
and should therefore be present in the Lagrangian.
In the case that we will eventually be
interested in -- the standard model -- the fermions carry gauge charges.
This forces the couplings $\lambda_{ij}$ and $m_{ij}$
to be block-diagonal, with mixing only between fields with identical
gauge quantum numbers. For simplicity we will set $\lambda_{ij}=\delta_{ij}$
in this paper, then $m_{ij}$ can be diagonalized with eigenvalues $m_i$.

Finding the massless four-dimensional fields is completely analogous
to the single fermion case of the last section. Each 5-d fermion $\Psi_i$
gives rise to a single 4-d left chiral fermion.
Again, the wave functions in the 5th coordinate are
Gaussian, but they are now centered around the zeros of $\Phi - m_i$.
In the SHO approximation this is at $x_5^i= m_i/2 \mu^2$. 
Thus, at energies well below $\mu$ the five-dimensional action above
describes a set of non-interacting four dimensional chiral fermions
localized at different 4-d ``slices'' in the 5th dimension.
Note that while the overall position of the massless fermions in the
$x_5$-direction is a dynamical variable (the location of the zero of $\Phi$),
the relative positions of the various fermions are fixed
by the $m_i$. Thus even when we turn on interactions
between the massless fields, the relative distances which control the
size of coupling constants in the effective 4-d theory stay fixed.

We now exhibit the field content of the 5-d theory which 
can reproduce the chiral spectrum of the 4-d SM as localized zero modes.
First note that by choosing all $\lambda$'s positive we have localized
only left handed chiral Weyl spinors. That implies that we will
construct the SM using only left handed spinors, the right handed
fields are represented by their charge conjugates
$\overline{\psi}^c$. Then the SM arises simply by choosing
5-d Dirac spinors $(Q,U^c,D^c,L,E^c)$ transforming 
like the left-handed SM Weyl fermions $(q,u^c,d^c,l,e^c)$.

We also briefly mention how we imagine confining gauge fields
to a (3+1)-dimensional wall. A field-theoretic mechanism
for localizing gauge fields was proposed by Dvali and Shifman
and was later extended and applied 
in \cite{ADD} (see also \cite{AHS}).
The idea is to arrange for the gauge group to confine outside
the wall; the flux lines of any electric sources turned
on inside the wall will then be repelled by the confining regions
outside and forced to propagate only inside the wall.
This traps a massless gauge field on the wall.
Since the gauge field is prevented to enter the confined region,
the thickness $L$ of the wall acts effectively as the size of the extra 
dimensions in which the gauge fields can propagate. Notice that in
a picture like this, the gauge couplings will exhibit power law
running above the scale $L^{-1}$, and so the scenario of \cite{DDG} for
gauge coupling unification may be implemented, without the 
presence of any new dimensions beyond the large gravitational dimensions. 

\section{Exponentially small 4-d couplings}
In this section we present two examples of applications for
our central result: exponentially small couplings
from small wave function overlaps of fields which are separated
in the fifth dimension. The two examples we consider are
SM Yukawa couplings and proton decay. Since our exponential suppression 
factors dominate any power suppression we will not keep track of the
various powers of scales which arise from matching 5-d to 4-d Lagrangians.

\subsection{Yukawa couplings}
In this section we apply our mechanism to generating hierarchical
Yukawa couplings in four dimensions.
Concentrating on only one generation 
and the lepton sector for the moment, we start with
the five-dimensional fermion fields with action

\beq
\CS = \int {\rm d}^5x\, \bar{L}[i\!\dsl_5 + \Phi(x_5)] L
      + \bar{E}^c[i\!\dsl_5+\Phi(x_5)-m] E^c \,  +\kappa H  L^{T} C_5 E^c .
\label{double5}
\eeq
where $C_5$ was defined in Eq. (\ref{C5}).
As discussed in the previous sections, we find
a left-handed massless fermions $l$ from
$L$ localized at $x_5=0$ and $ e^c$ from $E^c$
localized at $x_5=r \equiv m/(2 \mu^2)$. For simplicity, we will assume
that the Higgs is delocalized inside the wall.
We now determine what effective four-dimensional interactions between
the light fields results from the Yukawa coupling in eq. (\ref{double5}).
To this end we expand $L$ and $E^c$ as in eq.
(\ref{expansion}) and replace the Higgs field $H$ by its lowest
Kaluza-Klein mode which has an $x_5$-independent wave function.
We obtain for the Yukawa coupling

\beq
\CS_{Yuk} = \int {\rm d}^4{\bf x}\,\, \kappa\, h({\bf x}) l({\bf x}) e^c
({\bf x})\ \int {\rm d}x_5\ \phi_l(x_5)\ \phi_{e^c}(x_5)\ .
\label{yuk4}
\eeq
Here $\phi_l(x_5)$ and $\phi_{e^c}(x_5)$ are the zero-mode wave functions
for the lepton doublet and singlet respectively. $\phi_l$ is a Gaussian
centered at $x_5=0$ whereas $\phi_{e^c}$ is centered at
$x_5=r$. The
overlap of Gaussians is itself a Gaussian and we find
\beq
\int {\rm d}x_5\ \phi_l(x_5)\ \phi_{e^c}(x_5) = \frac{\sqrt{2}\mu}{\sqrt{\pi}}
\int {\rm d}x_5\ e^{-\mu^2x_5^2} e^{-\mu^2(x_5-r)^2}
= e^{-\mu^2r^2/2}\ .
\eeq

\begin{figure}[ht]
\centering
\epsfxsize=4.5in\epsfbox{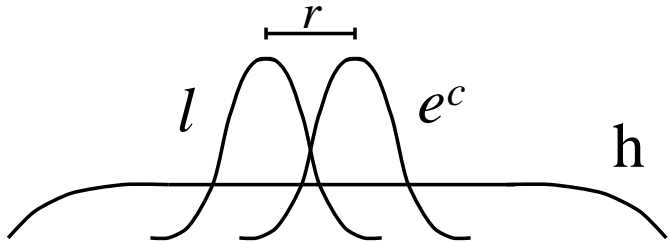}
\smallskip
\parbox{5in}{\caption{\it Yukawa coupling: the Gaussian wave functions of
the fermions $l$ and $e^c$ overlap only in an exponentially small
region, suppressing the effective Yukawa coupling exponentially.
}}
\end{figure}

This result is in agreement with the intuitive expectation from
Figure 2. Any coupling between the two chiral fermions is necessarily
exponentially suppressed because the two fields are separated in
space. The coupling is then proportional to the exponentially small
overlap of the wave functions.
 
Note that we did not impose any chiral symmetries in the fundamental
theory to obtain this result: the coupling $\kappa$ can violate the electron 
chiral symmetry by $O(1)$.  Even with chiral symmetry maximally broken
in the fundamental theory, we obtain an approximate chiral
symmetry in the low energy, 4-d effective theory.

\subsection{Long live the proton}

Proton decay places a very stringent constraint on most extensions
of the standard model. Unless a symmetry can be imposed to forbid
either baryon or lepton number violation, proton decay forces the
scale of new physics to be extremely high.
In particular one might be tempted to conclude that proton decay
kills all attempts to lower the fundamental Planck scale $M_*$
significantly beneath the
GUT scale, unless continuous or discrete gauge symmetries are invoked.
We now show that these no-go theorems are very elegantly evaded by
separating wave functions in the extra dimensions.
Consider for simplicity a one-generation model in five dimensions 
where the standard model fermions are again localized in the $x_5$
direction by coupling the five-dimensional fields to the domain wall
scalar $\Phi$. Assume that all quark fields are localized near
$x_5=0$ whereas the leptons are near $x_5=r$ as depicted schematically in
Figure 1. We allow the five-dimensional theory to violate both
baryon number and lepton number maximally, and we assume that we can
parameterize this violation by local operators\footnote{Non-local
operators which result from integrating out massive bulk
fields are discussed in the next subsection.}.
Then we can expect the following dangerous looking five-dimensional baryon
and lepton number violating operators
\beq
\CS \sim \int {\rm d}^5x\,  \frac{(Q^{T} C_5 L)^{\dagger} (U^{c T}
C_5 D^c)}{M_*^3}
\label{proton5}
\eeq
To obtain the corresponding four-dimensional proton decay operator we
simply replace the five-dimensional fields by the zero mode fields and
calculate the wave function overlap in $x_5$. The result is
\beq
\CS \sim \int {\rm d}^4{\bf x}\,\, \delta \times \frac{(q l)^{\dagger}
(u^c d^c)}{M_*^2}
\label{proton4}
\eeq
where
\beq
\delta \sim \int {\rm d}x_5\, \left[ e^{-\mu^2x_5^2}\right]^3
e^{-\mu^2(x_5-r)^2}\sim e^{-3/4 \mu^2r^2} \ .
\label{protonoverlap}
\eeq
Already for a separation of $\mu r =10$ we obtain $\delta\sim10^{-33}$
which renders these operators completely safe even for $M_* \sim 1$ TeV. 
Thus we imagine a picture where quarks and leptons are localized near
opposite ends of the wall so that $r \sim L$.
Once again, even if baryon and lepton number are maximally broken
in the 5-d theory
at short distances, the coupling generated in the 4-d
theory is exponentially suppressed and can
be harmless.  

In the following subsection, we present an alternate way of
understanding the suppression of proton
decay which also shows that corrections to this picture,
either coming from quantum loops or exchange of new degrees
of freedom, can be harmless.

\subsection{Long live the proton, again}
There is an alternative way of understanding the $e^{-(\mu r)^2}$
suppression which is physically transparent and
shows that all radiative corrections are also suppressed by the same
exponential factor. Even though it applies equally well to the case
of Yukawa couplings we will only describe the analysis for proton decay
here.
In order to decay the proton using the local $QQQL$ interaction, the
quarks and leptons must propagate into the bulk of the wall,
away from the points where they are massless (see Fig 4). 
\begin{figure}[ht]
\centering
\epsfxsize=4.0in\epsfbox{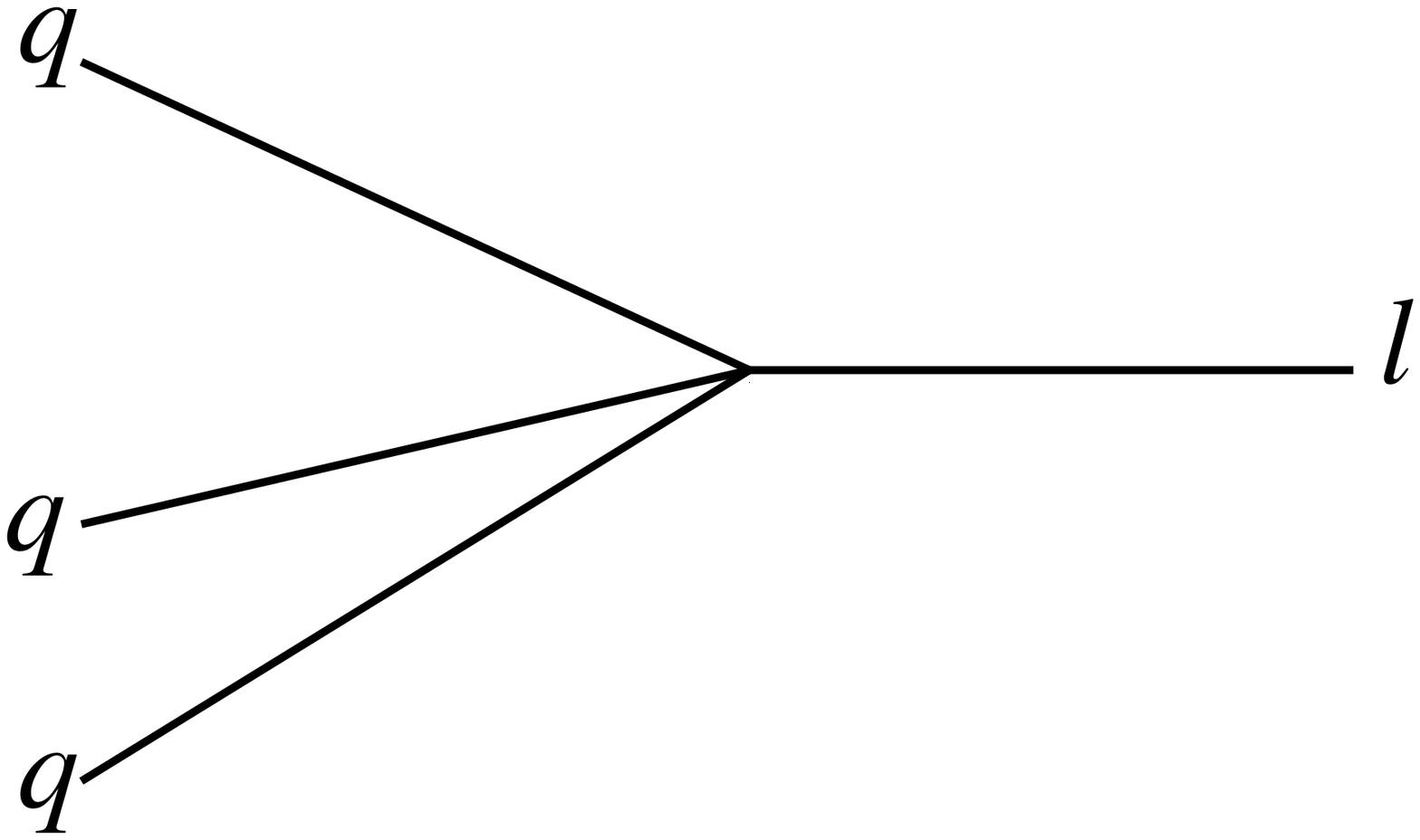}
\smallskip
\parbox{5in}{\caption{\it Tree-level proton decay diagram,
drawn in position space for the fifth dimension.
The quarks stuck at one end of the wall
and the lepton stuck at the other end 
propagate to some interior point s where they interact
via the higher-dimensional QQQL operator.
The ``free'' propagator to go to a point in the bulk is
given by the value of the (Gaussian) zero-mode wave function
at that point.
}}
\end{figure}
Because e.g.
the quarks are getting more massive as they move into the bulk, the
propagator from the plane where they live into the bulk is suppressed.
Intuitively, for each slice between $x_5$ and $x_5+ \delta x_5$, a Yukawa 
propagator $e^{-m(x_5) \delta x_5}$ must be paid. Therefore, the propagator to
reach a final point $x_*$ is proportional to 
\beq
\prod_{\mbox{slices}} e^{- m(x_5) \delta x_5} = e^{-\int^{x_*} m(x_5) dx_5} =
e^{-\mu^2 x_*^2} 
\eeq
for $m(x_5) = 2 \mu^2 x_5$. This is exactly the wave function for the zero mode
evaluated at $x_*$, as is intuitively expected 
and can also be seen more formally. In order to evaluate the tree level
diagram of Figure 4, we have to integrate over the interaction
vertex, yielding for the coefficient of the proton decay operator
\beq
\delta \sim \int dx_5 \left(\phi_q(x_5)\right)^3 \phi_l(x_5)\ ,
\eeq
precisely reproducing the result from our earlier ``overlap of
wave functions" picture.
This approach also makes clear why higher order corrections do not
significantly change the result. 
Indeed, the most general diagram for proton decay takes the form of
Figure 5: the effect of all interactions are encoded in
a modified propagator into the bulk and modified interaction vertex. The
modified propagator has the simple interpretation of 
being the wave function of the zero mode in the interacting theory. The
exact form of the modified vertex is unknown.
However, the vertex will still be point-like on scales of order $\mu$,
because all the interactions modifying it are mediated by particles
of mass $\mu$, which can only smear the vertex
on scales of order $\ge \mu$.
Since the propagators involved are needed at distances
$L \sim 10 \mu^{-1}$, the vertex in the Figure 5 is still 
effectively point-like, and so the picture of the suppression of proton
decay through exponentially small wave function overlaps persists,
if we replace the 
(free) Gaussian zero-mode wave functions by the true interacting ones.
\begin{figure}[ht]
\centering
\epsfxsize=4.0in\epsfbox{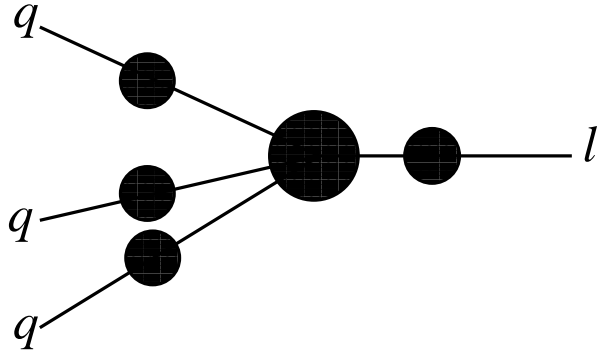}
\smallskip
\parbox{5in}{\caption{\it General proton decay diagram including
higher order effects.
The blobs on propagators denotes the all-order propagators,
and the blob on the vertex 
denotes the corrected vertex. The corrected propagator is
nothing other than the corrected
zero-mode wave function. The corrected vertex is still
local on scales larger than the width of the 
fermion wave function. This diagram is therefore well-approximated
by the overlap between the corrected
wave functions of the quark and lepton zero modes, which gives
the enormous suppression of proton decay.}}
\end{figure}

So far we have considered proton decay operators induced by
short-distance physics above the cutoff $M_*$; but what about effects
coming from integrating out fields possibly lighter than $M_*$?
In particular, we may worry that while the separation of quarks and
leptons suppresses higher-dimensional operators linking them,
operators involving only quarks on one side and violating baryon number,
or leptons on the other side violating lepton number, are not suppressed.
If a light field of mass $m$ freely propagates inside the wall, this 
may induce operators violating both $B$ and $L$ suppressed only by
$e^{-mL}$ (see Figure 6). However, in order to specifically induce 
proton decay, this light field would have to be fermionic.
In particular, no gauge or Higgs boson exchanges can
ever give rise to proton decay. If we make the single assumption that
all delocalized fermions have masses of order the cutoff $M_*$,
then their exchange can at most give $e^{-M_*L}$ contributions
which are comparable to the $e^{-(\mu L)^2}$ effects we have considered.
\begin{figure}[ht]
\centering
\epsfxsize=4.0in\epsfbox{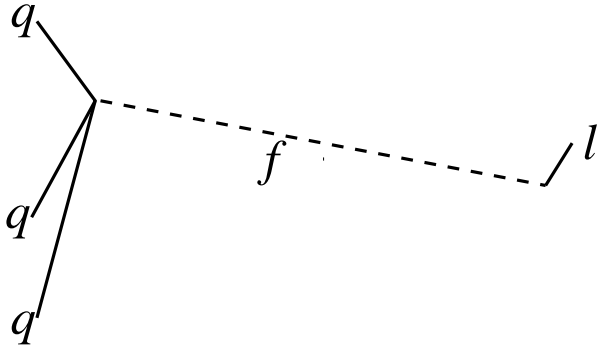}
\smallskip
\parbox{5in}{\caption{\it The exponential suppression of proton
decay through small wave function
overlaps can be avoided if there is a delocalized light fermion $f$ with
baryon and lepton number violating couplings as seen above. This
amplitude is only suppressed by the Yukawa factor $e^{-m_f L}$. In order
to avoid too large a rate for proton decay without imposing any symmetries,
we must postulate that there are no delocalized fermions lighter than $M_*$.}}
\end{figure}
Note that this argument implies that grand unification at
a scale as low as $\mu$ or $M_*$ does not lead to rapid proton
decay, as long as there are no delocalized fermionic fields with masses
below $M_*$. We cannot resist the temptation to speculate that the
same vacuum expectation values (VEVs) which break the GUT symmetry
near the scale $M_*$ may also be responsible for the separation
of the SM fermions in the
5'th dimension. For example the $m_{ij}$ of Eq. (\ref{multi5})
could stem from the VEV of a GUT symmetry breaking field which points
in the (B-L) direction. Then the SM fermions would be split according to
their baryon and lepton numbers. A VEV in the hypercharge direction
would arrange the fermions according to their hypercharge. 

We have seen that without imposing any symmetries on the
underlying theory, proton decay can be adequately suppressed
if quarks and leptons are stuck at different points in extra
dimensions. One might then wonder what happened to the general 
lore that black-hole/wormhole effects violate all non-gauged
symmetries and are therefore dangerous. In evaluating this argument,
we have to recall that it was Planck-scale sized wormholes
giving the supposedly $O(1)$ symmetry violating effects. Since these
have a mass above the cutoff, all their effects can be encoded
in terms of local operators suppressed by $M_p \ge M_*$; and indeed, we 
presumed that such ``dangerous" operators were really present
in the theory. However, their effects are harmless because the quarks 
and leptons are stuck at different points, yet have to be dragged
close to each other for the dangerous operators to be operative.
Of course, in the effective theory at distances longer than $L$,
the quarks and leptons look like they are on top of each other, so the
above suppression mechanism does not seem to apply.
However, only wormholes larger than $L$ are admissible in this
effective theory, and any effect they induce will be exponentially
suppressed by their action
\beq
e^{-S},\quad\;\; S \sim \int d^{4+n}x\, M_p^{(2+n)}\, {\cal R}
\sim (M_p L)^{(2+n)}
\eeq
which is a far larger suppression than the effects we have
computed $\sim e^{-(\mu L)^2} \sim e^{-M_*L}$.
The largest possible effect which might arise from wormholes would
come from long and skinny wormholes which stretch from the proton
to the lepton; but even these are completely safe as their action is
at least $\sim e^{-(M_pL)}$.

\subsection{Neutrino masses}
Separating quarks and leptons at different points in extra dimensions can
easily suppress proton decay without imposing any symmetries
on the high-energy theory. On the other hand, 
as already mentioned, operators violating baryon and lepton
number need not be suppressed. In fact, in the absence of any
symmetries in the high-energy theory, a Majorana neutrino mass operator
\beq
{\cal S} \sim \int d^5 x \frac{L^{T} C_5 L H^* H^*}{M_*^2}
\eeq
turns into an unsuppressed Majorana mass term for the 4-d zero mode
\beq
{\cal S} \sim \int d^4 {\bf x} \frac{l l h^* h^*}{M_*}
\eeq
since there is no small overlap between $l$'s wave function
with itself. There are a number of ways of resolving this 
problem; we will just mention the obvious strategy of adding
a right-handed neutrino and gauging $(B-L)$. Of course
$(B-L)$ must be broken in such a way as to not allow large
Majorana masses after breaking. In our framework this would be
most naturally achieved with a $(B-L)$ breaking VEV which is localized
within the wall but at some distance from the lepton field $l$ so
that the Majorana neutrino mass is exponentially suppressed
\footnote{$(B-L)$ could also be broken everywhere within the wall
if a discrete subgroup remains preserved, or it could be broken
on a distant wall if $(B-L)$ is gauged in the large bulk where gravity
propagates \cite{ADD2,AD}. For another approach to neutrino
masses see \cite{mor}.}. In addition one would also get small
Dirac neutrino masses, with the tiny Yukawa couplings originating from the
overlap between right- and left- handed neutrino wave functions.

\subsection{Summary of scales}
Let us close by giving an account of the various scales we are
now imagining.
Recall that the at the edge of the wall, the fermion mass $\langle\Phi\rangle$
is $\sim \mu^2 L \sim 10 \mu$, and must not be larger
than the ultraviolet cutoff $M_*$. In fact we will take them to be
comparable. Therefore, we have three scales in the problem:
the UV cutoff $M_*$, $\mu$ and the wall thickness scale $L^{-1}$,
with magnitudes related roughly as
\beq
M_* \sim 10 \mu \sim 100 L^{-1}
\eeq
Since we cannot push $L^{-1}$ significantly below $\sim$ TeV, the
fundamental scale $M_*$ is bounded below by $\sim 100$ TeV.
This is actually desirable from another point of view: in the absence
of flavor symmetries, it is difficult to protect against flavor changing
neutral currents without
pushing the scale of higher-dimension operators to $\sim 100$ TeV. 

Notice that even though the theory becomes effectively 5-dimensional
above $L^{-1}$, the theory is perturbative up to the UV cutoff $M_*$.
From the 4-dimensional viewpoint, we have
$N_{KK} \sim (M_* L)/2 \pi \sim 10-100$ gauge and Higgs
field KK modes, and so the effective
expansion parameter is 
\beq
\frac{h_4^2}{16 \pi^2} \times N_{KK} \sim O(1)
\eeq
where $h_4$ is a generic low-energy gauge coupling or top
Yukawa coupling. From the higher-dimensional point of view, the 
theory is on the edge of being strongly coupled at the UV cutoff $M_*$.

Finally, we wish to give a rough idea of the sort of suppressions
which are generated by our $e^{-(c \mu L)^2}$ size effects,
with $c=1/2$ for Yukawa couplings and $c=3/4$ for proton decay:

\vskip .1in
\begin{tabular}{c|cccccccccc}
$\mu r$          & 0 & 1 & 2     & 3     & 4     &  5  & 7 && 10 \\ \hline
exp$(-c \mu^2r^2)$ &1&1&$10^{-1}$&$10^{-2}$&$10^{-4}$&$10^{-6}$&$10^{-11}$&&
$10^{-33}$ \\  \hline 
& & $\lambda_t$ & & $\cdots$ &  & $\lambda_e$ &  $\cdots$ &
& $\Gamma_{proton}$ \\
\end{tabular}
\vskip .1in
It is attractive that for $\mu r$ just ranging between
$1 - 10$, we can get appropriate sizes for everything from the top
Yukawa coupling $\lambda_t$ (for $\mu r \sim 1$),
to the electron Yukawa $\lambda_e$
($\mu r \sim 5$), to sufficient suppression for proton decay
($\mu r \sim 10$).

\section{Cartography with gauge fields}
While the SM fermion fields are stuck at different points in the
extra dimension, the gauge fields are totally delocalized, and
we expect that we can probe the locations of the fermions using the
gauge fields. Cartography of the SM fermions with gauge fields
will become an experimental science to be performed at the LHC or
NLC if the wall thickness is as large as $\sim 1$ TeV$^{-1}$.

To see how this works explicitly, consider a toy 
example with two 4-d chiral fermions $\psi_1, \psi_2$, 
transforming identically under a gauge group $G$, but stuck to 
different points $s_1,s_2$ in the extra dimensions.
At distances larger than the width of their wavefunction
in the extra dimensions, the dynamics that localizes
the fermions is irrelevant; we can approximate their wavefunctions
as delta functions or, what is the same, fix them to live
on a 3 dimensional wall, while gauge fields freely propagate in the bulk.
The effective coupling to gauge fields is fixed by gauge invariance to be
\beq
{\cal S} \sim \int d^4 {\bf x}\ \bar{\psi}_1 \bar{\sigma}^{\mu} T^a
\psi_1 A^a_{\mu}({\bf x},s_1) + 
\bar{\psi}_2 \bar{\sigma}^{\mu} \psi_2 A_{\mu}({\bf x},s_2)
\label{effc}
\eeq
Let us Fourier expand $A_\mu$ as 
\beq
A_\mu({\bf x},x_5) = {A_\mu^{(0)}({\bf x})\over \sqrt{2}} +
\sum_{n=1}^\infty A_\mu^{(n)}({\bf x}) \mbox{cos}(k_n x_5) +
\sum_{n=1}^\infty B_\mu^{(n)}({\bf x}) \mbox{sin}(k_n x_5),
\eeq
where $k_n=\frac{2 \pi}{L} n$. Here $A_{\mu}^{(0)}$ corresponds to
the massless 4-d gauge field,
and the $A_\mu^{(n)}$ and $B_\mu^{(n)}$ are KK excitations with masses $k_n$.
Inserting this expansion into Eq. (\ref{effc}) we find
the effective couplings of the tower of 4-d KK gauge fields to the fermions.
Note that since $A_{\mu}^{(0)}$ has a flat wave function in the extra
dimensions, it has the same coupling to both fermions as required by
4-dimensional gauge invariance. However the couplings of the fermions
to the massive KK states, $A$ and $B$, are proportional to cosines
and sines from the wave functions of the KK states at the locations
of the fermions \footnote{Note that these couplings are valid
for the KK modes with 
wavelength long compared to the fermion localization width;
shorter wavelength KK modes can resolve the fermion wave function
and so the delta function approximation for the fermion wave function
is inadequate.}. 

\begin{figure}[ht]
\centering
\epsfxsize=4.5in\epsfbox{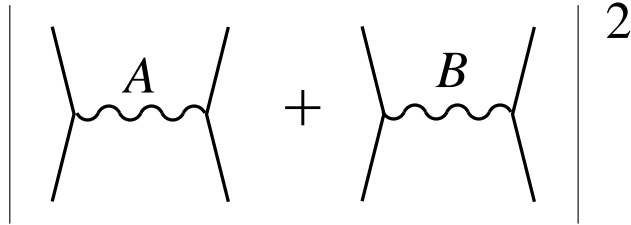}
\smallskip
\parbox{5in}{\caption{\it KK exchange diagrams.}}
\end{figure}

Suppose that we are sitting on the $s$-channel resonance 
for particle-antiparticle annihilation mediated by the $n$'th KK modes
of $A$ and $B$.
Then the relative cross section for $1^+ 1^- \to 1^+ 1^-$ (and
$2^+ 2^- \to 2^+ 2^-$) as calculated from the diagrams in Figure 7
will be different from $1^+ 1^- \to 2^+ 2^-$:
\bea
\sigma(1^+ 1^- \to A^{(n)}, B^{(n)} \to 1^+ 1^-) & \equiv & \sigma_n \nonumber \\
\sigma(2^+ 2^- \to A^{(n)}, B^{(n)} \to 2^+ 2^-) & = & \sigma_n \nonumber \\
\sigma(1^+ 1^- \to A^{(n)}, B^{(n)} \to 2^+ 2^-) & = & \sigma_n\ 
\mbox{cos}^2(k_n(s_1 - s_2)) \ ,
\eea
and this can be used to gain information on the distance between
$\psi_1,\psi_2$ in the extra dimensions.

Actually, there are subtleties in this analysis associated with
the mechanism for localizing gauge fields.
For example, in the Dvali-Shifman mechanism,
confinement outside the wall forces specific boundary
conditions $F_{\mu 5} = 0$ at the edges of the wall \cite{AHS}.
This then enforces $\partial_5 A_{\mu}({\bf x},x_5) = 0$ at
$x_5=0,L$, and the $B^{(n)}$ are eliminated. In addition, the $A^{(n)}$ may
now be periodic or anti-periodic, thus $k_n=\frac{\pi}{L} n$.
The position dependence of KK gauge couplings raises the interesting
possibility that some of the KK excitations of various fields may be
leptophobic or baryophobic if the quarks or leptons sit at nodes of
their wave functions.
More generally, the cross sections above are modified to
\bea
\sigma(1^+ 1^- \to A^{(n)} \to 1^+ 1^-) & =&\sigma_n\ \mbox{cos}^4(k_n s_1) \\
\sigma(2^+ 2^- \to A^{(n)} \to 2^+ 2^-) & =&\sigma_n\ \mbox{cos}^4(k_n s_2) \\
\sigma(1^+ 1^- \to A^{(n)} \to 2^+ 2^-) & =&\sigma_n\ \mbox{cos}^2 (k_n s_1)
\ \mbox{cos}^2(k_n s_2)
\eea

There are clearly other interesting possibilities arising from the
non-universal couplings of SM fermions to the KK excitations. As one
example, in our scenario for suppressing proton decay by separating quark
and lepton wave functions, the non-standard coupling of the quarks and
leptons to the KK modes has an interesting impact on atomic parity
violation (APV). The latest experimental results \cite{apv} indicate that
the measured weak charge of the nucleus is {\it lower} than the SM
expectation by $\sim 2.5 \sigma$. If we had a conventional
Kaluza-Klein tower at $\sim$ TeV scale, with standard couplings to quarks
and leptons, this would {\it enhance} the SM contribution to APV. 
In our case, however, the situation can be different. If we impose the
$F_{\mu 5}=0$ boundary conditions as stated above, then the first
Kaluza-Klein excitation has the profile shown in Figure 8. Notice that the
product of the quark and lepton couplings to the first KK excitation 
has the opposite sign as in the SM, and gives a contribution to atomic
parity violation that moves in the right direction. The sign of this
effect is an inevitable consequence of our mechanism for suppressing
proton decay, and the correct magnitude can be obtained if the wall
thickness is $\sim 1$ TeV.

\begin{figure}[ht]
\centering
\epsfxsize=5.5in\epsfbox{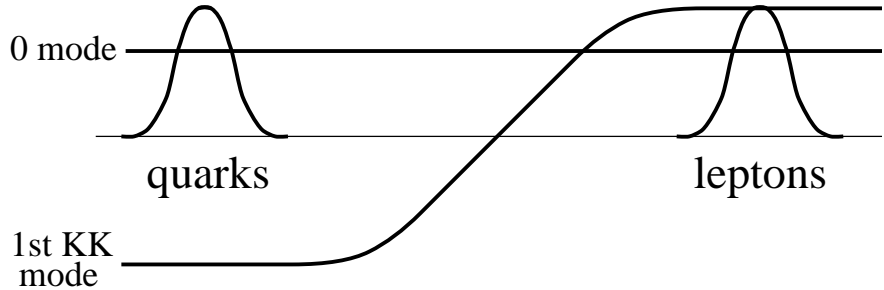}
\smallskip
\parbox{5in}{\caption{\it Since the wave functions of the usual SM gauge
fields (zero modes) are flat in the extra dimensions, they have
identical couplings to
the quarks and leptons stuck at opposite ends of the wall, as
required by gauge invariance. On the other hand, since the first KK
excitation has a non-trivial wave function, its couplings can be different.
In particular, the product of quark and lepton gauge couplings to the
first KK excitation has the opposite sign as for the SM.}}
\end{figure}

\section{Conclusions}
In this paper we have shown that approximate symmetries can arise in
a long-distance 
theory without any symmetry explanation in the underlying short-distance
theory.
Instead, even if symmetries are maximally broken at short distances,  
exponentially small couplings between different fields can result 
if they are ``stuck" at slightly different points in extra dimensions.
This opens 
a new arena for model-building, where a specific arrangement of
the fermions in extra dimensions, and not familiar
flavor symmetries, determine the fermion mass hierarchy.
Furthermore, proton decay can be elegantly disposed of, even in
theories with the fundamental cutoff close to the TeV scale, 
if quarks and leptons are separated from each other by a factor
of 10 larger than their size in the extra dimensions.
If the effective size of this extra dimension or equivalently our
wall thickness is close to the TeV scale, these ideas can be probed
at the LHC and NLC. The smoking gun for our mechanism for would be the 
detection non-universality in the coupling of SM fermions  
to the KK excitations of the SM gauge fields. A detailed analysis
of this non-universality could then be used to ``map" the locations of the 
fermions in the extra dimensions.
In closing we would like to emphasize that our mechanism of suppressing
couplings from non-trivial wave functions is generically operative in
higher dimensional theories with chiral fermions, as most such models
obtain chiral matter from modes stuck to a defect in the higher dimensions.
This defect may be a field theoretic domain wall in one extra dimension,
a cosmic string in two extra dimensions, or a D-brane or orbifold
fixed point\footnote{The $e^{-r^2}$ suppression of Yukawa couplings
between twisted sector fields at different orbifold fixed points has
been known for some time, see e.g. \cite{ibanez}.} in a string model.

\section{Acknowledgements} We thank Savas Dimopoulos, Lance Dixon, Gia Dvali,
David Kaplan, John March-Russell, Gino Mirabelli, Shmuel Nussinov,
Michael Peskin, Tom Rizzo and Eva Silverstein for discussions.
Our work is supported by the Department of Energy under contract
DE-AC03-76SF00515.
 
\def\pl#1#2#3{{\it Phys. Lett. }{\bf B#1~}(19#2)~#3}
\def\zp#1#2#3{{\it Z. Phys. }{\bf C#1~}(19#2)~#3}
\def\prl#1#2#3{{\it Phys. Rev. Lett. }{\bf #1~}(19#2)~#3}
\def\rmp#1#2#3{{\it Rev. Mod. Phys. }{\bf #1~}(19#2)~#3}
\def\prep#1#2#3{{\it Phys. Rep. }{\bf #1~}(19#2)~#3}
\def\pr#1#2#3{{\it Phys. Rev. }{\bf D#1~}(19#2)~#3}
\def\np#1#2#3{{\it Nucl. Phys. }{\bf B#1~}(19#2)~#3}
\def\mpl#1#2#3{{\it Mod. Phys. Lett. }{\bf #1~}(19#2)~#3}
\def\arnps#1#2#3{{\it Annu. Rev. Nucl. Part. Sci. }{\bf #1~}(19#2)~#3}
\def\sjnp#1#2#3{{\it Sov. J. Nucl. Phys. }{\bf #1~}(19#2)~#3}
\def\jetp#1#2#3{{\it JETP Lett. }{\bf #1~}(19#2)~#3}
\def\app#1#2#3{{\it Acta Phys. Polon. }{\bf #1~}(19#2)~#3}
\def\rnc#1#2#3{{\it Riv. Nuovo Cim. }{\bf #1~}(19#2)~#3}
\def\ap#1#2#3{{\it Ann. Phys. }{\bf #1~}(19#2)~#3}
\def\ptp#1#2#3{{\it Prog. Theor. Phys. }{\bf #1~}(19#2)~#3}

\end{document}